\documentclass{WileyMSP-template}

\usepackage{graphicx}%
\usepackage{multirow}%
\usepackage{amsmath,amssymb,amsfonts}%
\usepackage{amsthm}%
\usepackage{mathrsfs}%
\usepackage[title]{appendix}%
\usepackage{xcolor}%
\usepackage{textcomp}%
\usepackage{manyfoot}%
\usepackage{booktabs}%
\usepackage{algorithm}%
\usepackage{algorithmicx}%
\usepackage{algpseudocode}%
\usepackage{listings}%

\usepackage[bookmarks=true,colorlinks=true,linktocpage=true,backref=true]{hyperref}
\usepackage{endnotes}
\usepackage{hyperendnotes}

\begin{document}

\pagestyle{fancy}
\rhead{\includegraphics[width=2.5cm]{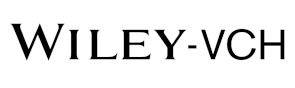}}

\title{Ergodic Concepts for a Self-Organizing Trivalent Spin Network: A Path to $(2+1)$-dimensional Black Hole Entropy}

\maketitle


\author{Christine C. Dantas}


\dedication{}

\begin{affiliations}
Divis\~ao de Astrof\'{\i}sica\\
Instituto Nacional de Pesquisas Espaciais\\
Av. dos Astronautas, 1758, Jardim da Granja\\
S\~ao Jos\'e dos Campos - 12227-010 - SP - Brazil\\
Email Address: christine.dantas@inpe.br\\
\end{affiliations}


\keywords{loop quantum gravity, spin network model, ergodic theory, self-organized critical phenomenon}

\begin{abstract}
 {We consider, from a dynamical systems point of view, a trivalent spin network model in Loop Quantum Gravity presenting self-organized criticality (SOC), arising from a spin propagation dynamics. We obtain a partition function for the domains of stability connecting gauge non-invariant avalanches, leading to an entropy formula for the asymptotic SOC state. The microscopic origin of this SOC entropy is therefore given by the excitation-relaxation spin dynamics in the avalanche cycle. The puncturing of TSN edges participating in the avalanche are counted in terms of an ensemble perimeter  
over the implicit avalanches. By identifying this perimeter with that of an isolated $(2+1)$-dim. black hole horizon, we conjecture that the SOC entropy reduces to the Bekenstein-Hawking perimeter-entropy law for the Ba\~nados, Teitelboim, and Zanelli (BTZ) black hole, by an appropriate adjustment of a potential function based on the thermodynamical formalism of Sinai, Ruelle and Bowen.}
\end{abstract}


\section{Introduction}\label{SEC_introduction}

{ {
\subsection{LQG and SOC: how does macroscopic spacetime arise?}}

One of the most important unsolved problems in fundamental physics is to establish a robust and testable theory of quantum gravity \cite{Car01}, \cite{Kie07}, \cite{Lol22}, \cite{Add22}, \cite{Kie23}.
{\sl Loop Quantum Gravity} (LQG), for example,  has achieved a considerable body of development \cite{Rov2000}, \cite{RovBOOK07}, \cite{ThiBOOK08}, \cite{RovVidBOOK20}. It is based on a background independent quantization of general relativity (GR), in which quantum states of geometry are conceptualized in terms of {\sl spin networks}. Hence, spacetime would be discrete at the Planck scale, $\ell_{\rm Planck} \equiv \sqrt{G\hbar/c^3} \sim 10^{-33} ~{\rm cm}$. 

\medskip

{One of the key challenges in LQG is describing the continuum limit of the theory, which corresponds to the smooth spacetime of GR. Several coarse-graining approaches are currently being explored, representing an active area of research (\cite{Asa22} and references therein). A particularly intriguing question is whether {\sl phase transitions} could lead to the emergence of a physically smooth spacetime \cite{Mar97}, \cite{Fel16}, possibly resulting from {\sl self-organized critical} (SOC) phenomena \cite{Bak87}, \cite{Bak88}, \cite{Dha90}, \cite{Marko14}, \cite{Wat16}. SOC, initially proposed by Bak et al. \cite{Bak87}, offers an explanation for power-law behavior (scale invariance) observed in certain systems without the need for fine-tuning the parameters. A well-known example of SOC is the ``sandpile'' model, where the mean surface of the pile forms an angle with the horizontal. When a grain of sand is randomly dropped onto the pile, it causes a disturbance that may result in a downward avalanche, depending on whether the angle exceeds a critical threshold. The system thus transports sand mass until it reaches a state of equilibrium between avalanches.}

\medskip

{An analogy can be drawn with condensed matter physics, where the phase transition of a material from an ordered state (solid) to a disordered state (liquid or gas) can be compared to the transition of spacetime from a quantized, discrete state to a continuous one.  In the context of SOC, the critical state is characterized by the absence of a characteristic scale, meaning that the same dynamics govern the system across different scales, whether small or large avalanches. The system remains perpetually on the cusp of a transition, with the potential for substantial changes even from minimal perturbations. These large-scale changes can be viewed as a mechanism through which spacetime might emerge macroscopically.}

\medskip

One of the first  investigations of SOC in the context of LQG was made in \cite{Bor99}, using ``frozen'' {\sl trivalent spin networks} (TSN) with a {\sl random edge} model, which, however,  did not produce a critical behavior.  Subsequently, Ansari and Smolin \cite{Ans08} revisited the question, with simulations defined on a {\sl random vertex} model, resulting in SOC. Other studies, also based on numerical simulations, confirmed SOC in different TSN models \cite{Che08}, \cite{Dan21-CQG}.  {Note that the term ``frozen'' should be understood in the sense that the TSN connectivity does not change over time. However, the model includes spin propagation rules on the edges of {\sl gauge non-invariant  (GNI)} vertices, which means that the spin labels on the edges of the network can evolve. While the connectivity of the TSN remains unchanged, the dynamics of the spin propagation indicate that the model has internal dynamical processes. And such processes are what in fact allows for the observation of self-organizing behavior.}

\medskip

 {As SOC investigations in LQG were based on TSN models, they imply reduced representations the quantum state of the gravitational field. In fact, the TSN is embedded on $2$-dimensional spacelike surface, therefore with vanishing volume quanta. However, the study of quantum gravity in $(2+1)$-dimensions has shown to be of great interest as a conceptual model, given the technical difficulties in finding a complete quantum theory of space-time \cite{Car98}, \cite{Car2023}. A deeper understanding of SOC in dimensionally reduced models in LQG could offer insights into the potential relevance of SOC in the {full 4-dimensional Lorentzian case (spin foam models) \cite{Per2013}}, suggesting it as a viable mechanism for achieving the continuum limit of the theory.}

{ {
\subsection{The BTZ black hole: insights into the microscopic origin of entropy?}}

\medskip

 {What are the microscopic (i.e., quantum gravitational) excitation states that give rise to black hole entropy, and where are they located? Despite the proposal of several state-counting methods across various unrelated approaches, the question of how to derive black hole thermodynamics from a statistical mechanical source remains elusive \cite{Page2005}. To tackle this issue, the simplicity of dimensionally reduced gravity can provide valuable insights. For instance, the entropy of the $(2+1)$-dim. black hole of Ba\~nados, Teitelboim, and Zanelli (BTZ;  \cite{BTZ92} \cite{Ban93}) follows a completely analogous relation as the {\sl Bekenstein-Hawking area-entropy law} \cite{Haw72}, \cite{Bek73}, with the area being substituted by the  perimeter of the $1$-dim. event horizon  \cite{Car95}, \cite{Car1995PhRvD}:}

\begin{equation}
 {S_{\rm BTZ} = {2 \pi r_+\over 4 \hbar G};} \label{BTZ_Entropy}
\end{equation}

\noindent  {where $r_+$ is the horizon radius.}

\medskip

 {Given this key result, BTZ black holes can be used as a simplified model for black hole physics.  It is important to note that the BTZ metric is an exact solution of the dimensionally reduced Einstein's field equations with a negative cosmological constant (asymptotically anti-de Sitter space)  \cite{Car98}. Otherwise, no black hole solutions with event horizons exist for a zero cosmological constant in $(2+1)$-dim. In this sense, the BTZ black hole also plays a role in understanding the AdS/CFT (anti-de Sitter/conformal field theory) correspondence, providing potential insights into the holographic nature of gravity (e.g., \cite{Car05} and references therein).}

\medskip

 {A crucial point is how to count the microstates that contribute to the entropy of a black hole \cite{Str98}, \cite{Bir98}, \cite{Car1998CQG}. In this regard, the proposed derivations of the BTZ black hole entropy do not trivially generalize to the {full 4-dimensional Lorentzian} case \cite{Car07}.  From the fundamentals of statistical mechanics, entropy is obtained from the logarithm of the state density, as a function of energy. Two prevalent methods provide a measure of entropy: (i) a directly counting process (i.e., starting from a vacuum state, enumerate the various ways to add excitations to achieve a given energy); (ii) an indirect process (i.e., obtaining a partition function to derive the density of states). However, these methods failed to provide a conclusive answer concerning the nature of the relevant degrees of freedom and where these excitations reside (e.g., on a boundary, to be identified with the event horizon, or at infinite, described by a
conformal field theory with a central charge?)  \cite{Car1998CQG}, \cite{Car2023}.}

\medskip

 {In the particular case of LQG, derivations for the Bekenstein-Hawking area-entropy law were obtained from the microcanonical and canonical ensembles, e.g.,  \cite{Rov96}, \cite{Kra97}, \cite{Loc12}, \cite{Per17}. The BTZ entropy has also been derived in LQG from similar principles, e.g., \cite{Fro13}.
However, despite these results and derivations based in other quantum gravity approaches,  conceptual aspects related to the microscopic origin of the (whether physical or BTZ) black hole entropy remain unsolved \cite{Car2023}. A complete quantum gravity theory is not yet available but still several independent methods converge to the same entropy, despite difficult issues concerning the nature of microscopic states \cite{Car05}, \cite{Car07}.  For one side, this indicates some cross-validation of the essential aspects of black hole thermodynamics, but for the other side, it might indicate an oversimplification of the underlying microscopic physics, reducing the capacity for theory differentiation and testing. }

\medskip

{ {
\subsection{Motivations and outline of this work}}

 {{\sl \underline{Ergodic concepts for a SOC TSN?}} Powerful mathematical frameworks for the analysis of many-body systems have been systematically developed since the last century. These include  {\sl dynamical systems theory} \cite{Bar12}, encompassing {\sl ergodic theory} \cite{Via16} and the {\sl thermodynamical formalism} (TDF) of Sinai, Ruelle and Bowen \cite{Sin72},  \cite{Rue78}, \cite{Boe08}. These methods have also been used in the analysis of SOC.  For instance, the Zhang model \cite{Zha89} was investigated using such theoretical tools, leading to insights into SOC  dynamics \cite{Bla99}, \cite{Ces01}, \cite{Ces04}. We here refer to these tools collectively as ``ergodic concepts'',  given that the principle of ergodicity and the results derived from it are fundamental to their use.}

\medskip

 {Motivated by the promising application of these mathematical frameworks to the Zhang model, we apply them in the study of the TNS SOC model by Ansari and Smolin  \cite{Ans08}.  The most important observation is the fact that the standard techniques of statistical mechanics cannot be directly applied to SOC systems \cite{Ces04}. This is because the steady state is achieved through a specific non-Hamiltonian microscopic dynamics, leading to the absence of a ``natural'' Gibbs distribution or free energy in these systems. TDF has been successfully applied to the Zhang model by constructing equivalents of finite-volume Gibbs measures, in which the Hamiltonian was replaced with a specific potential form. In this way, generating functions for the avalanche distributions could be written down explicitly, serving as formal equivalents to partition functions and free energy in statistical mechanics.}

\medskip

 {We follow closely Refs. \cite{Bla99}, \cite{Ces01} and \cite{Ces04} for several concepts throughout our analysis. We focus on connecting  microscopic dynamics to macroscopic observables, such as entropy. This quantity is well-defined within ergodic theory, provided that certain conditions, such as hyperbolicity and ergodicity, are satisfied. We present propositions concerning these and other aspects with explicit variables and parameters outlined. Some of these propositions, with the necessary adaptations, are assumed without proof, as they closely follow the analogous results from the Zhang model, where proofs are available. Others are not trivial and their justification is here outlined in some detail. This preliminary work focuses on establishing a proof of concept for applying ergodic concepts in quantum gravity models, starting with a relatively simple application. The necessary adaptations have been made, and detailed proofs will be pursued in a future study.} 

\medskip

 {{\sl \underline{Do spin avalanches lead to the BTZ  entropy?}} By successfully determining a suitable form for the SOC TSN partition function over a set of legal sequences of avalanches, we can readily compute the corresponding entropy. Therefore, in our framework, the microscopic origin of the SOC TSN entropy arises from the excitation-relaxation spin dynamics within the avalanche cycle. 
These degrees of freedom do not reside in a pre-established boundary in the TSN, but arise dynamically from the domains of stability connecting gauge non-invariant avalanches.}

\medskip

 {We associate the puncturing of TSN edges participating in the avalanche in terms of an ensemble perimeter over the domains of stability. Only in the final step of the derivation, this perimeter is identified with that of an isolated $(2 + 1)$-dim. black hole horizon. We establish a connection between the resulting SOC TSN entropy and the entropy of a BTZ black hole (Eq. \ref{BTZ_Entropy}) through a potential function in the formalism. Both entropies are shown to converge to the same form.  }

\medskip

 {Our paper is organized as follows. In Sec. 2, we introduce the mathematical modeling for the propagation rules for the frozen TSN, its configuration, excitation/relaxation maps and avalanches. In Sec. 3, we introduce the SOC TSN model in the formal language of dynamical systems. In Sec. 4, we state the main propositions underlying our main conjecture, specially related to stability maps, ergodicity, potential and partition functions. In Sec. 5 we present our main conjecture for the derivation of the BTZ black hole entropy from state-counting arising from GNI avalanches. In Sec. 6, we present our conclusions. We use units in which $c = 1$ {and $k_{\rm B}$ = 1}.}

\medskip

{ {
\section{TSN configuration, excitation, relaxation and avalanches}\label{SEC_methodology}}

{ {
\subsection{The frozen TSN model}\label{SEC_model}}

The TSN is a labelled graph consisting of vertices, each with $3$ oriented edges connecting neighboring vertices. Edges are labelled by the irreducible representation of the SU($2$) group (in terms of spins, $j$, or colors, given by $ {\bar{c}}= 2j$). The dual space of the TSN is a (planar) triangulation of a  ($2$-dim.) region of space.  The {\it gauge-invariance constraint} (GIC) \cite{Ans08}, \cite{RovBOOK07}, \cite{ThiBOOK08} ,\cite{RovVidBOOK20} on a vertex with edge colors $\mathbf{c} = (a,b,c)$ is the set of conditions (the dual triangle inequalities):
\begin{equation}
[GIC]~~~~
\mathcal {C}_1:  a+b \geq c; ~~~
\mathcal {C}_2:  a+c \geq b; ~~~
\mathcal {C}_3:  b+c \geq a; ~~~
\mathcal {C}_4:  a+b+c = {\rm even}.\label{EQN-CONSTR}
\end{equation}

\medskip

The {\sl propagation rules} for the frozen TSN model are:

\medskip

1. {{\sl Initialization of the TSN:} Set the {\it time iterate}, $\tau =1$. Initial edge colors are random even integers ($\neq 0$), with GIC (Eq. \ref{EQN-CONSTR}) satisfied in the whole TSN. }

\medskip

2. {{\sl Excitation:} Choose a random (activating) vertex, $v_{\alpha}$, and subtract a fixed color value  {$\Delta \bar{c}$} from all its edges. If at least one of the edges is $0$, re-start this procedure by choosing another vertex.  The quantity  {$\Delta \bar{c}$} represents an external driving disturbance.}

\medskip

3. {{\sl Relaxation:} Sweep the TSN (using a fixed ordering; e.g., left to right and top to bottom) for testing the GIC. One sweep is denoted a {\sl relaxation cycle}. For each analysed vertex:} {\sl (i)} if {\sl gauge invariance} (GI) is satisfied, proceed to the next vertex without changing its state; {\sl (ii)} if the vertex is {\sl gauge non-invariant} (GNI), add simultaneously to each of its edges the color disturbance  {$\Delta \bar{c}$}, and proceed the sweeping to the next vertex.

\medskip

4. {At the end of a relaxation cycle, if the whole TSN satisfies the GIC, then follow the next step, otherwise repeat a new relaxation cycle (go to item 3).}

\medskip

5. {{\sl Time iterate update:} set $\tau = \tau + 1$. Go to item 2 and repeat the procedure.}

\medskip

{ {\subsection{Matrices and mappings}\label{SEC_matrices}}

We follow closely Refs. \cite{Bla99}, \cite{Ces01} and \cite{Ces04}. Main differences concern the type of system (the Zhang model being a $d$-dimensional lattice, versus the TSN being a planar graph), the critical parameter (the energy ``stored'' in nodes in the Zhang model, versus the color ``stored'' in the TSN edges), and the disturbance rules, which are different in both cases.  We consider a finite subgraph, $\Lambda^{\rm TSN}$, with $n$ vertices, { {and with assigned coordinates given by} $m = \{1, \dots, n \}$, which is taken as the sweeping order.  Let the {\sl vertex set} of a TSN  be: $\mathcal{V}(\Lambda^{\rm TSN}) = \{ v_1, \dots v_n \}$, and its {\sl edge set} be: $\mathcal{E}(\Lambda^{\rm TSN}) = \{ e_{i,j} \}$, with edge $e_{i,j}  (= e_{j,i})$ connecting the adjacent vertices $v_i$ and $v_j$. Consider $n \times n$ (symmetric) {\sl weighted adjacency matrices} \cite{Gro01}, given by: 

\medskip

\noindent $\bullet$ The {\sl configuration matrix}, characterizing univocally a given configuration of edge colors in $\Lambda^{\rm TSN}$, where $\mathsf{c}(e_{i,j})$ is a function that extracts the color value of the edge $e_{i,j}$:
\begin{equation}
\mathsf{A}(\Lambda^{\rm TSN}) \equiv
\left\{ 
\begin{array}{ll}
A_{ij} =  \mathsf{c}(e_{i,j})      & \mbox{,  if there is an edge} ~e_{i,j}, \\ 
A_{ij} = 0       & \mbox{, if there is no edge}.\\
A_{ii} = 0,      &  
\end{array}
\right. \label{MATRIXA}
\end{equation}
\noindent $\bullet$ The {\sl disturbance  matrix}:
\begin{equation}
\mathsf{D}_m(\Lambda^{\rm TSN}) \equiv
\left\{ 
\begin{array}{ll}
D_{ij} =  Z_m {\Delta \bar{c}}    & \mbox{, if there is an edge} ~e_{i,j}, \\ 
D_{ij} = 0      & \mbox{, if there is no edge}, \\
D_{ii} = 0,      &  
\end{array}
\right. \label{MATRIXD}
\end{equation}
\noindent where $Z_m = 0$ if $e_{i,j} = e_{m,j}$ belongs to a GI vertex $v_m$; otherwise, $Z_m = 1$ for a GNI $v_m$. Set $Z_m = -1$, if the edge $e_{m,j}$ belongs to the {\sl current activation vertex}, $v_{\alpha}(\tau)$.

\medskip

Let $\mathcal{M}_{\rm color} = (\mathbb{Z}_{+})^n$ be a $n$-dim. discrete space of edge color values. A state $\mathsf{A}$ in the form of Eq. (\ref{MATRIXA}) can be univocally represented by a point in this space, and if all its elements obey the GIC, it will belong to a {\sl stable subspace} $\mathcal{M} \subset \mathcal{M}_{\rm color}$, whereas any GNI vertex in $\mathsf{A}$ will place it in an {\sl unstable subspace} $\overline{\mathcal{M}} \subset \mathcal{M}_{\rm color}$.  

\medskip

The {\sl excitation map}, $\mathsf{E}_{\alpha}:{\mathcal{M}} \rightarrow \{ \mathcal{M} ~{\rm or}~ \overline{\mathcal{M}} \}$, is easily implemented as an entrywise matrix sum of the form: $\mathsf{E}_{\alpha}(\mathsf{A}) = \mathsf{D}_{\alpha} + \mathsf{A} = \overline{\mathsf{A}}$, which may lead to a stable or unstable state. A relaxation cycle (sweep) is a recursive process of the form: $\mathsf{R}_{\rm cycle}(\overline{\mathsf{A}}): ~~\overline{\mathsf{A}}_{(m+1)} = \mathsf{D}_m + \overline{\mathsf{A}}_{(m)},  ~~~{\rm for~} m = \{1, \dots , n-1 \}. $ If there is no instability after excitation, all $Z_m$'s are zero, and $\mathsf{R}_{\rm cycle}(\overline{\mathsf{A}}) =  \mathsf{I} \overline{\mathsf{A}}  = \overline{\mathsf{A}} \subset \mathcal{M}$ (the identity map). We assume that this recursive sweep is implemented in a {\sl relaxation map} of the form:
$\mathsf{F}: \overline{\mathcal{M}} \rightarrow \mathcal{M}$:
\begin{equation}
\mathsf{F}_r (\overline{\mathsf{A}}) = 
\mathsf{A}^{\prime}, ~~~ r = \{ 1, 2, \dots, r_{{\rm max}}(\tau) \}, \label{MAPF}
\end{equation}
\noindent where the superscript $r$ above symbolically denotes the action of $r_{{\rm max}}$ sweeps, occurring until GI is achieved in the whole TSN, with state $\mathsf{A}^{\prime}(\tau) \subset \mathcal{M}$.

\medskip

{ {
\subsection{Avalanches}\label{SEC_avalanches}}

We define an {\sl avalanche}, at any given $\tau$, as the set of vertices subject to relaxation. We define the {\it size of the avalanche} as the total number of vertices participating in an avalanche, with recurring GNI vertices in the same avalanche  repeatedly counted. We label each avalanche with the double index $(\alpha, \beta)$, in which $\beta$ labels the different avalanches that occur {\sl starting at the same vertex}, $v_{\alpha}$.  The relaxation map, $\mathsf{F}: \overline{\mathcal{M}} \rightarrow \mathcal{M}$ (Eq. \ref{MAPF}), connects all unstable states achievable in an avalanche which starts from any point in $\mathcal{M}$ that has been excited.
Hence, $\mathcal{M}$ is an extended space of ``GI pyramids'', in the terminology given in \cite{Ans08} for the single vertex space. $\mathcal{M}$ is delimited by a (stable) boundary, $\partial \mathcal{M}$, which is a higher-dimensional analogue of the ``sheets of flatness'' \cite{Ans08}, given by the critical conditions:
$a+b = c; ~~ a+c = b; ~~ b+c = a$.

\medskip

\medskip

{ {
\section{The TSN as a SOC dynamical system}\label{SEC_dynamical}}

{ {
\subsection{Stability, domains of continuity, evolution of the TSN and asymptote} \label{SEC_domains}}

We combine the previous maps into a {\sl stability map}, $\mathsf{T}_{\alpha}:\mathcal{M} \rightarrow \mathcal{M}$, which associates a stable configuration to another stable configuration, with an implicit avalanche in between:
\begin{equation}
{\rm (Stability)}: ~~ \mathsf{T}_{\alpha}(\mathsf{A}) = 
\mathsf{F}_{{r}_{\alpha}} (\mathsf{D}_{\alpha} + \mathsf{A}),~~~ \mathsf{A} \in \mathcal{M}, \label{ER}
\end{equation}
\noindent where  {${r}_{\alpha}$} labels the number, $r$, of relaxation cycles involved in producing the avalanche from $v_{\alpha}$. Note that this map inherits the sweeping recursive procedure, and it has an adjacency matrix representation.  {The structure of $\mathsf{T}_{\alpha}$ indicates that it can be composed of maps, $\mathsf{T}_{(\alpha, \beta)}$, with  {\sl domains of continuity}, $\mathcal{M}_{(\alpha,\beta)}$:}
 {
\begin{equation}
\mathsf{T}_{\alpha} \rightarrow \mathsf{T}_{(\alpha_1, \beta_1)}
\circ \dots \circ \mathsf{T}_{(\alpha_1, \beta_{{\rm max}})} 
\circ \dots \circ \mathsf{T}_{(\alpha_{\rm max}, \beta_1)} 
\circ \dots \circ \mathsf{T}_{(\alpha_{\rm max}, \beta_{{\rm max}})},
\label{COMPOSITION}
\end{equation}}
\noindent so that there is a one-to-one correspondence between the set of avalanches, $\{(\alpha, \beta)\}$, and the set of maps $\{ \mathsf{T}_{(\alpha,\beta)} \}$.  {At the neighborhood of $\mathsf{A}$,  the discrete differential at $\mathsf{A}$ is:}
 {
\begin{equation}
{\rm D} \mathsf{T}_{\alpha}(\mathsf{A}) \approx  \mathsf{I} +
 \mathsf{T}_{\alpha}(\mathsf{A}), \label{DOP}
\end{equation}}
\noindent {where we implicitly assume in this linear approximation a minimum ${r}_{\alpha}$ generating a stable configuration.}

\medskip

Consider the excitation of random vertices in the TSN as a discrete stochastic process. Namely, the vertex $v_{\alpha} \in \mathcal{V}(\Lambda^{\rm TSN})$ is chosen with probability $p(v_{\alpha}) = 1/n$.  The sample space is given by the set $\Sigma^{+}_{\Lambda^{\rm TSN}}$ of right infinite  {\sl activation sequences}, $\mathbf{a}$:
$\Sigma^{+}_{\Lambda^{\rm TSN}} = \{ \mathbf{a} = (i_1, i_2, \dots i_k, \dots ): i_k \in \Lambda^{\rm TSN}\}$, where $i_k$ is the $k$th. activation vertex in the sequence $\mathbf{a}$. The activation dynamics is given by the {\sl left Bernoulli shift operator}: 
\begin{equation}
\sigma(i_k) = i_{k+1}, \label{EQ-SIGMA}
\end{equation}
\noindent so that, e.g., $\sigma \mathbf{a} = \sigma (i_1, i_2, \dots) = i_2,i_3\dots$. We denote the first element of $\mathbf{a}$ as $a_1$, and $[a_1]$ denotes the set of activation sequences whose first digit is $a_1$.
Let $\Omega = \Sigma^{+}_{\Lambda^{\rm TSN}} \times \mathcal{M}$ be the {\sl extended phase space}, with the associated dynamical vector in this space being: 
$\widehat{\mathbf{X}} \equiv (\mathbf{a},\mathsf{A}) \in \Omega$, with projections: 
$\pi^u (\mathbf{a},\mathsf{A}) = \mathbf{a};
\pi^s (\mathbf{a},\mathsf{A}) = \mathsf{A}$. The induced partition of $\Omega$ is: $\mathcal{P} = \{ \mathcal{P}_{(\alpha, \beta)} = [\alpha] \times \mathcal{M}_{(\alpha, \beta)}\}$.  The dynamics in the extended phase-space is then traced in terms of {\sl return maps} \cite{Bla99}. We define the {\sl evolution of frozen SOC TSN dynamical system} as the map $\mathcal{F}: \Omega \rightarrow \Omega$ (Fig. \ref{FIG-DYN}):
\begin{equation}
\mathcal{F}(\widehat{\mathbf{X}})  \equiv 
\left ( \sigma\mathbf{a} , \mathsf{T}_{a_1}(\mathsf{A}) \right ). \label{DYNSYS}
\end{equation}

\begin{figure}
\centering
\includegraphics[width=0.6 \linewidth]{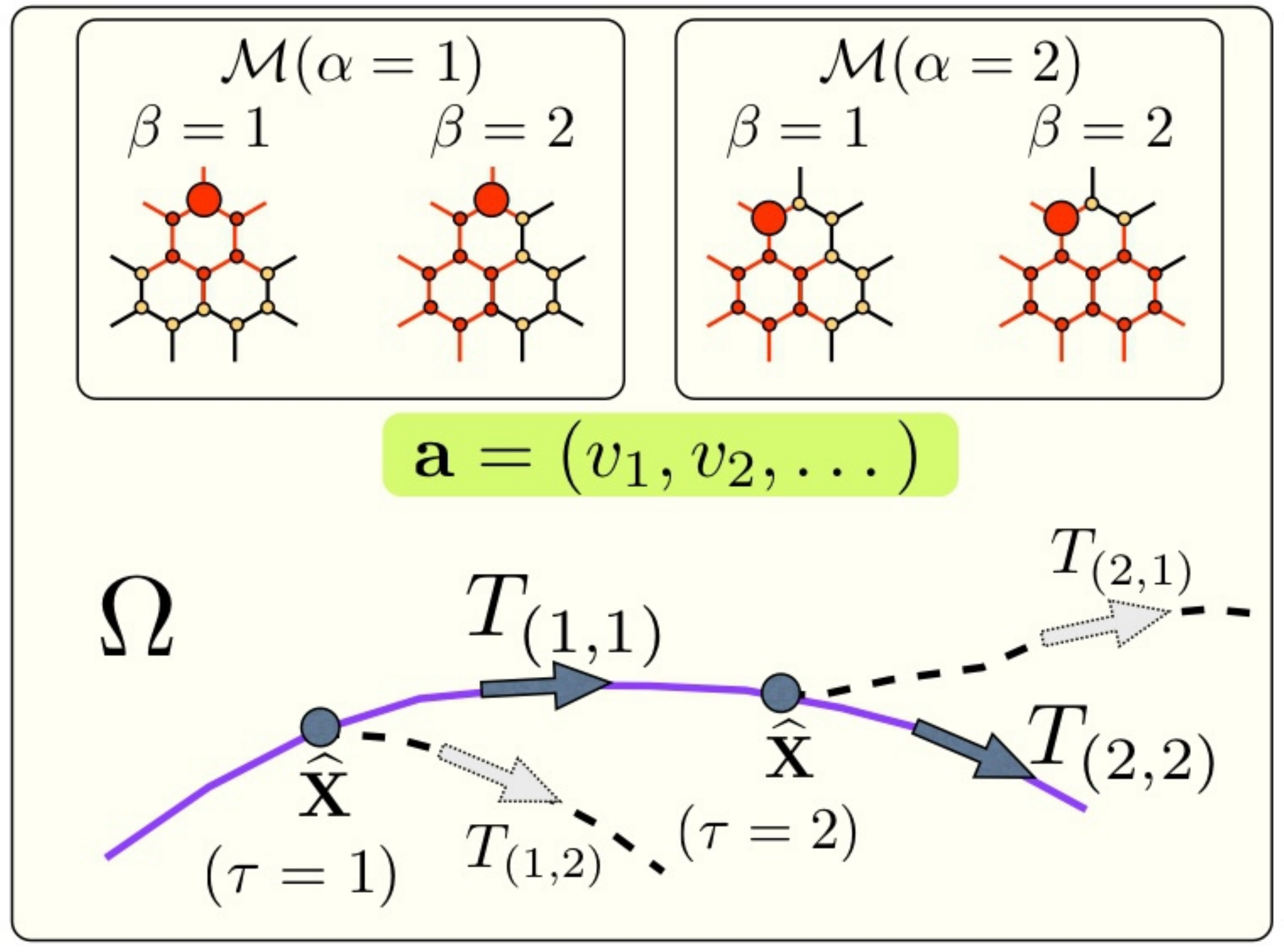} 
\caption{A schematic illustration of the dynamical system $\Omega$, for only a small set of possibilities. The two upper panels show the domains $\mathcal{M}_{(\alpha, \beta)}$, with the larger circle representing the activating vertex $v_{\alpha}$, resulting in only $2$ avalanches (labeled by $\beta = 1,2$), which are indicated by red in the TSN's. The configuration $\widehat{\mathbf{X}}(\tau)$ follows an orbit in the space $\Omega$, with activating sequence $\mathbf{a}$. The paths in dashed lines are allowed paths for the domains of continuity, but are not followed by the given  $\widehat{\mathbf{X}}(\tau)$.
\label{FIG-DYN}}
\end{figure}

\medskip

{ {
\subsection{The attractive asymptote}\label{SEC_asymptote}}

{The propagation rules generally lead to an increase in the average color of the TSN as a function of $\tau$ (see, for example, Fig. 7 in \cite{Ans08} and Fig. 3 in \cite{Dan21-CQG}, where the average color increase follows an approximately exponential trend with $\tau$). The underlying SOC dynamics do not result in a fixed-point attractor as the system evolves in $\mathcal{M}$, but rather converge statistically towards an attractive {\sl asymptote} (for similar cases, see \cite{Boe97} and \cite{Nor02}; for a detailed algorithm on computing space asymptotes that may simplify the analysis of complex dynamical systems, refer to \cite{Bla15}). We employ a scaling procedure to achieve a stationary description similar to that presented in \cite{Nor02}, which is justified here by the fact that the GIC (Eq. \ref{EQN-CONSTR}) remains invariant if all colors are rescaled by the same factor. In this context, we assume that $\tau$ {\sl begins counting after an initial transient period}, during which edge colors are {\sl implicitly rescaled} by $f_c(\tau) \propto \exp(-k \tau)$, where $k$ is a constant.}

\hspace{4cm}
$~$

{ {
\section{Propositions: stability maps, ergodicity, potential and partition functions  }\label{SEC_propositions}}

{{\sl \underline{Proposition 1}.} At a time iterate $\tau$,  {we define the norm of the discrete operator acting on the stability map, in its decomposition into  $\mathsf{T}_{(\alpha, \beta)}$, as (c.f. Eqs. \ref{ER},  \ref{COMPOSITION} and \ref{DOP})}:
\begin{equation}
\|  {\rm D} \mathsf{T}_{(\alpha, \beta)} \|  \equiv
\det[{\rm D} \mathsf{T}_{(\alpha, \beta)}] \approx \epsilon_{\tau}^{\mathcal{S}_{\rm h}(\alpha,\beta)}, 
~~~~ \epsilon_{\tau} \equiv \epsilon(\tau) \equiv    {{r}_{\alpha}(\tau) \Delta \bar{c}}, \label{DETCOLL}
\end{equation}
\noindent where $\mathcal{S}_{\rm h}(\alpha, \beta)$  gives the avalanche size of all vertices belonging to a complete {\sl hexagonal pattern} of the TSN, so $\mathcal{S}_{\rm h}(\alpha, \beta)$  is always a multiple of $12$ (to be explained next). 

\hspace{1cm}

\noindent {\sl --- Justification of Prop. 1.}  Let $p_{\rm subg}$ be the number of {\sl spanning subgraphs} of $\Lambda^{\rm TSN}$, then:
\begin{equation}
\det({\rm D} \mathsf{T}_{(\alpha, \beta)}) \approx \det(\mathsf{I}) +
\det(\mathsf{T}_{(\alpha, \beta)}) \approx 
\sum_{p=1}^{p_{\rm subg}} (-1)^{e_p} 2^{c_p} 
\prod_{y_{\lambda} \in L_p} (y_{\lambda})^2 
\prod_{y_{\xi} \in M_p} (y_{\xi}), \label{DET}
\end{equation}
\noindent where the {\sl adjacency determinant} above is  defined in the context of {\sl graph theory}, in which we used Eqs. (5) and (6) of \cite{Har62}. $L_p$ is the set of subgraphs consisting of two vertices and the edge joining them, and $M_p$ the set of remaining subgraphs, each of which is a cycle. The $y$'s  correspond to non-zero elements of the (adjacency) matrix representation of $\mathsf{T}_{(\alpha,\beta)}$, where the subscript $\lambda$ (or $\xi$) runs for all edges in the corresponding subgraphs $L_p$ (or $M_p$). 
In Fig. \ref{FIG1}(a), we show the {\sl smallest TSN with one hexagonal pattern}, where we consider that all its $12$ vertices underwent one avalanche ($\mathcal{S}_{\rm h} = 12 \equiv 1 \times \mathcal{S}_{12}$). The labels in this figure are such to facilitate the notation.  The subgraphs ($L_1, L_2, L_3, M_1$) are shown in Fig. \ref{FIG1}(b). From Eq. (\ref{DET}), and  {${r}_{\alpha} = 1$}, we find:  
\begin{align} 
\det(\mathsf{T}_{\rm small~ TSN}) &= 
\prod_{\lambda = 1}^6 (y_{\lambda})^2 
- \prod_{\lambda = 7,9,11} (y_{\lambda})^2 
- \prod_{\lambda = 8,10,12} (y_{\lambda})^2 
- 2\prod_{\xi = 7}^{12} y_{\xi}   \nonumber \\
  &=  { (\Delta \bar{c})^{12} -  (\Delta \bar{c})^{6} -  (\Delta \bar{c})^{6} -2  (\Delta \bar{c})^{6} 
 \approx (\Delta \bar{c})^{12} \equiv (\Delta \bar{c})^{1 \times \mathcal{S}_{12}}. }
  \label{DETPATT}
\end{align}
\noindent Note the important feature of the above determinant acting on such a graph (with one hexagonal pattern): terms are non-zero only if {\sl all} vertices in the hexagonal pattern participate in the relaxation. A partial relaxation of it would not count for the determinant above. To obtain the general expression for the determinant acting on a larger TSN (Eq. \ref{DETCOLL}), with $n_{\mathcal{S}}$ hexagonal avalanches, we assume   submaps, $\mathsf{T}_k$, of the form of $\mathsf{T}_{\rm small~ TSN}$ (Eq. \ref{DETPATT}), with avalanche sizes as multiples of $\mathcal{S}_{12}$, say: 
 {
$$\det(\mathsf{T}) \approx 
\det (\mathsf{T}_1)\det (\mathsf{T}_2) \dots \det(\mathsf{T}_{n_{\mathcal{S}}}) =  
(\Delta \bar{c})^{n_1 \times \mathcal{S}_{12}}
(\Delta \bar{c})^{n_2 \times \mathcal{S}_{12}}\dots
(\Delta \bar{c})^{n_{n_{\mathcal{S}}} \times \mathcal{S}_{12}}
=  (\Delta \bar{c})^{\mathcal{S}_{\rm h}},$$
\noindent with ${r}_{\alpha}$ entering in the $y$'s.}

\begin{figure}
\centering
\includegraphics[width=0.46 \linewidth]{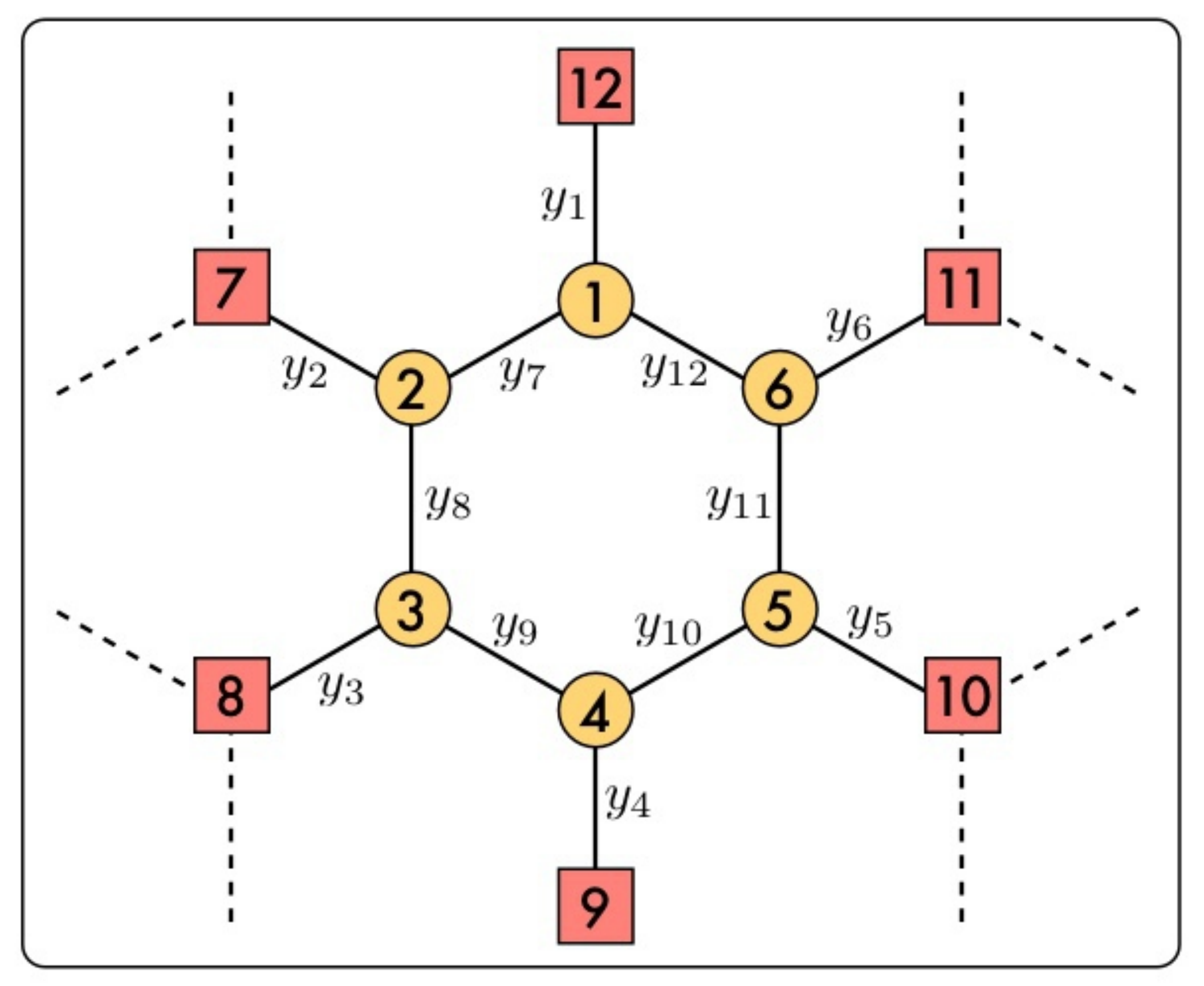} 
\includegraphics[width=0.40 \linewidth]{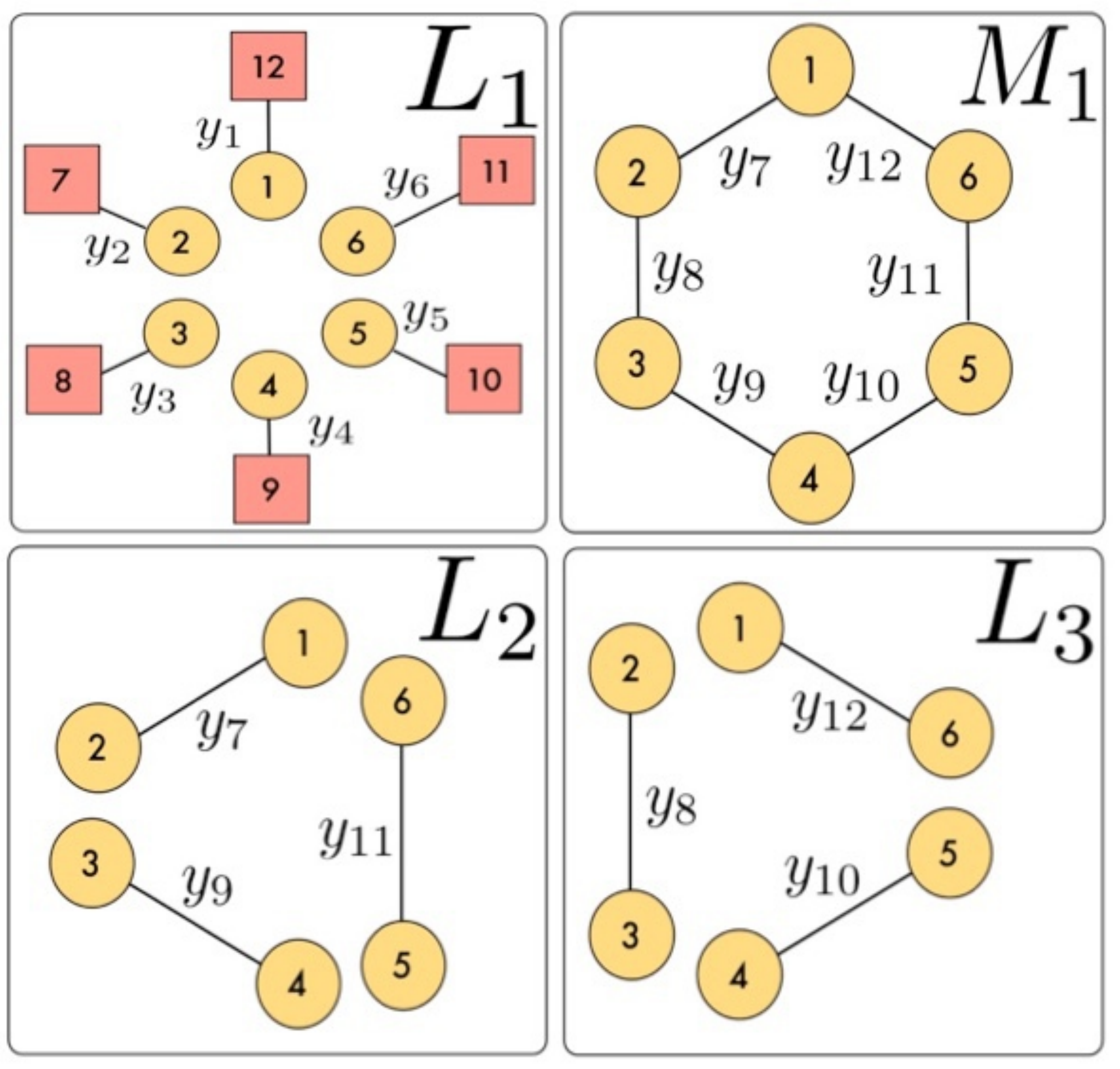} \\
\hspace{0.2cm}(a) \hspace{4.5cm}(b)
\caption{{\sl (a)} The smallest TSN with one hexagonal pattern (vertices in squares are at the boundary of the TSN, with only one edge being considered in each). {\sl (b)} Subgraphs $L_1, L_2, L_3, M_1$. \label{FIG1}}
\end{figure}

\hspace{1cm}

{{\sl \underline{Proposition 2}.}  The dynamical system (Eq. \ref{DYNSYS}) shares  {\sl piecewise hyperbolic} properties \cite{Via16} (in a weak sense, due to the finiteness of  $\Lambda^{\rm TSN}$), so that we can make use of several established results available for such systems. Let a compact set be $Q \subset \Omega$, representing different configurations of $\widehat{\mathbf{X}}$. $Q$ is $\mathcal{F}$-invariant, i.e., $\mathcal{F}^{-1}Q = Q$ ($\mathcal{F}^{-1}$ is defined from the inverse of all maps involved in the definition of $\mathcal{F}$). We define $\mathsf{T}_{\widehat{\mathbf{X}}}\Omega$ as the {\sl tangent space} at ${\widehat{\mathbf{X}}} \in \Omega$, with the {\sl tangent map} at ${\widehat{\mathbf{X}}}$ being a properly defined discrete differential, ${\rm D}\mathcal{F}_{\widehat{\mathbf{X}}}$ \cite{Bla99}. We assume the Oseledec theorem is valid (c.f. details in \cite{Ces01}), so that there exists a decomposition of $\mathsf{T}_{\widehat{\mathbf{X}}}\Omega$ in terms of {\sl stable} $\mathcal{E}^s$ and  {\sl unstable} $\mathcal{E}^u$ subspaces at the point $\widehat{\mathbf{X}}$: 
$
\mathsf{T}_{\widehat{\mathbf{X}}}\Omega = 
\mathcal{E}^s(\widehat{\mathbf{X}}) \oplus
\mathcal{E}^u(\widehat{\mathbf{X}})
$, with a hierarchy of nested subspaces:  
$\mathcal{E}^s(\widehat{\mathbf{X}}) = 
\mathcal{E}_1(\widehat{\mathbf{X}}) \supset
\mathcal{E}_2(\widehat{\mathbf{X}}) \supset
\dots \supset
\mathcal{E}_{\ell}(\widehat{\mathbf{X}})
$, and a hierarchy of {\sl Lyapunov exponents}:
\begin{equation}
\lambda_l (\widehat{\mathbf{X}}) \equiv 
\lim_{\tau \rightarrow \infty} 
{1\over \tau} \log 
\| {\rm D}\mathcal{F}^{~(\tau)}_{\widehat{\mathbf{X}}} v \| ~ ~ ~ l = \{ 1, \dots n\}, 
v \in \mathcal{E}_l(\widehat{\mathbf{X}}) \setminus\mathcal{E}_{l+1}(\widehat{\mathbf{X}}), \label{LYA}
\end{equation}
\noindent which are negative in $\mathcal{E}^s$, and positive in $\mathcal{E}^u$ {, with only one exponent, representing the ``entropy of the excitation dynamics'':}
\begin{equation}
 {
\lambda_0 = \log(n) \equiv S_{\rm exc}. \label{LAMBDA0} }
\end{equation}
\noindent Hence the exponents characterize the decay or expansion of {the} norm of a small perturbation ($v$, in the tangent space at $\widehat{\mathbf{X}}$) under the action of the infinite differential map. 
 
\hspace{1cm}

{{\sl \underline{Proposition 3}.}  {We consider the family of potential functions, $\phi$, in the TDF \cite{Ces04}, in terms of the Lyapunov exponents, here identified as (c.f. Eqs. \ref{LYA} and \ref{LAMBDA0}):}
  {
\begin{equation}
\phi_{q,\eta} = -q S_{\rm exc} + \eta \log \{ \det [{\rm D}\mathsf{T}_{(\alpha, \beta)} ] \},
\end{equation}}
\noindent   {where $q$, $\eta$, are free parameters. From Eq. (\ref{DETCOLL}), we write our {\sl TNS SOC potential function}:}
\begin{equation}
\phi(\alpha, \beta, \tau)_{q,\eta} \equiv 
-qS_{\rm exc} + \eta \mathbf{s}, \label {POT}
\end{equation}
\noindent  {where}
\begin{equation}
\mathbf{s}  \equiv \mathbf{s}(\alpha, \beta, \tau)   \equiv \log(\epsilon_{\tau})\mathcal{S}_{\rm h}(\alpha,\beta). \label{ATAU}
\end{equation}

\medskip

 {Note that since $\mathbf{s}$ relates to the avalanche size over several relaxation cycles, a perimeter measure can be established as the puncturing of TSN edges participating in an avalanche over a $1$-dim. surface. The latter naturally arises from an avalanche $(\alpha,\beta)$ belonging to the domains of continuity in the $n$-dim. space of edge colors,  $\mathcal{M}_{(\alpha, \beta)}$, which are delimited by the boundary, $\partial \mathcal{M}$, a higher-dimensional analogue of the ``sheets of flatness'' \cite{Ans08} (c.f. Secs. \ref{SEC_matrices}, \ref{SEC_avalanches}, and \ref{SEC_domains}). Note that such a boundary at some time iterate $\tau$ corresponds to a $1$-dim. perimeter from the point of view of the underlying  planar TSN.}

\medskip

 {We identify the parameter $\eta$ as a {\sl conjugate perimeter}, whereas $q$ is a pure number (c.f. Eq. \ref{LAMBDA0}). We can set the conjugate perimeter normalized by $q$ as:}
 {
\begin{equation}
\theta = {\eta \over q}  ~~~~~~ {\rm (conjugate ~perimeter)}. \label{thetadef}
\end{equation}}

 {We absorb the ``excitation entropy'' in the potential function (c.f. Eqs. \ref{LAMBDA0}, \ref{POT}), as:}

 {
\begin{equation}
{1 \over q}\phi(\alpha, \beta, \tau)_{q,\eta} +S_{\rm exc} \equiv \bar{\phi} = \theta \mathbf{s}, \label {POTnew}
\end{equation}}
  
\hspace{1cm}

{{\sl \underline{Proposition 4}.}  Consider a {\sl finite partition function} from TDF \cite{Ces04}:
 {
\begin{equation}
Z_T(\bar{\phi}) = \sum_{\{ \rm seq \}} \exp \left \{ 
\sum_{\tau=1}^T\bar{\phi}({\rm orbit}) \right \}, \label{POTTDF}
\end{equation}
\noindent where we simplified the notation so that the first summation refers to a set of legal sequences of avalanches $(\alpha,\beta)$ shifted along the domains of continuity,  $\mathcal{M}_{(\alpha, \beta)}$, leading to an orbit in $\Omega$ (Eq. \ref{DYNSYS}), for the potential function, $\bar{\phi}$, Eq. (\ref{POTnew}).}

\medskip

\noindent  {At time iterate, $\tau$, let $N_{\rm h}$ be distinguishable hexagonal avalanches, $n_{\rm h}$ of which have size $\mathcal{S}_{\rm h}(\alpha,\beta)$, then, using Eqs. (\ref{POTnew}) and (\ref{POTTDF}), we define a partition function representing an (asymptotic) statistical ensemble over  $\mathcal{M}_{(\alpha, \beta)}$: 
\begin{equation}
 Z_T(\theta, N_{\rm h}) \propto \sum_{\{ n_{\rm h} \}} {N_{\rm h} !\over \prod_{\tau} n_{\rm h}!}
  \prod_{\tau = 1}^T 
   \left \{
   \exp \left ( -\theta \mathbf{s} \right )  
   \right \} 
   ^{n_{\rm h}}
 \propto \left \{
 \sum_{\tau=1}^T   
 \exp \left ( -\theta \mathbf{s}\right )  
 \right \}^{N_{\rm h}}
 \propto z^{N_{\rm h}}(\theta). \label{ZT2}   
\end{equation}
\noindent where we used the the binomial theorem in the last equality, and dropped the dependence of $N_{\rm h}$  and $n_{\rm h}$  on $\tau$ for clarity.}

\hspace{1cm}

{{\sl \underline{Proposition 5}.}  We assume {\sl ergodicity}  \cite{Bar12}, \cite{Via16}, namely, an almost everywhere equality between the {\sl time iterate average}, $\overline f$, and the {\sl ensemble average}, $< f >$, so that for any function $f$ on $\Omega$:
\begin{equation}
\overline f(\widehat{\mathbf{X}}) = \lim_{T \rightarrow \infty} {1 \over T}
\sum_{\tau = 1}^T f[ \mathcal{F}^{(\tau)}(\widehat{\mathbf{X}}) ] 
= Z_T^{-1} \sum_Q e^{-\theta \mathbf{s} } f(\widehat{\mathbf{X}})  = < f (\widehat{\mathbf{X}}) >, ~\forall \widehat{\mathbf{X}} \in Q. \label{ERGO}
\end{equation}

\hspace{1cm}

 {{\sl \underline{Proposition 6}.}  {We define a function on edge colors, $\mathcal{C}_{r_+}(\widehat{\mathbf{X}})$, that gives the equivalent circumference of radius $r_+$, resulting from the puncturing of TSN edges (in the state $\widehat{\mathbf{X}}$ at a given $\tau$) into a perimeter of the TSN. This perimeter corresponds, in the $n$-dim. space of edge colors, to a boundary, $\partial \mathcal{M}$, of a given domain of continuity,  $\mathcal{M}_{(\alpha, \beta)}$, for an avalanche $(\alpha,\beta)$ (c.f. Proposition 3). We define $\mathcal{C} (Q)$ as the ensemble average of this perimeter within a region $Q$ of $\Omega$, of given by:
\begin{equation}
\mathcal{C}_Q \equiv \mathcal{C} (Q)= Z_T^{-1} \sum_Q e^{-\theta \mathbf{s}} \mathcal{C}_{r_+}(\widehat{\mathbf{X}}) , ~~~~~\forall \widehat{\mathbf{X}} \in Q. \label{ENSAV}
\end{equation}
 }

{ {
\section{Conjecture: the BTZ black hole entropy from GNI avalanches}\label{SEC_conjecture}}

{\sl \underline{Conjecture}.}   {There is a {\sl perimeter-entropy law} in the form:}
 {
\begin{equation}
S_Q =\theta(u_0)  {\mathcal{C}_Q}, \label{ENTROPYperim}
\end{equation}
\noindent where
\begin{equation}
\theta(u_0) \equiv \frac{\eta(u_0)}{q}, \quad u_0 = \text{constant},
\end{equation}}

\noindent  {where the conjugate perimeter, $\theta(u_0)$, determines the scaling of the potential function, Eqs. (\ref{thetadef}) and (\ref{POTnew}), and $\mathcal{C}_Q$ is the ensemble perimeter, Eq. (\ref{ENSAV}). By an appropriate adjustment of the parameters $\eta(u_0)$ and $q$,  $S_Q$ can be made formally equal to the entropy of the BTZ black hole, $S_{\rm BTZ}$ (Eq. \ref{BTZ_Entropy}), by taking the ensemble perimeter as $\mathcal{C}_Q = 2 \pi r_+$, for some equivalent circumference radius $r_+$ in the ensemble.}

\hspace{1cm}

\noindent {\sl --- Justification.}  {We follow closely the derivation given in Ref. \cite{Loc12}, adapted to the reduced dimensionality of the problem. Hence, the following variables/parameters in that paper are translated to  (namely, \cite{Loc12} $\Leftrightarrow$ this paper):}
 {
\begin{equation}
\begin{aligned}
 A     \quad \text{(area)} \quad  & \Leftrightarrow \quad \mathcal{C}_Q \quad \text{(perimeter)} \\
 \beta \quad \text{(conjugate area)} \quad & \Leftrightarrow \quad \theta \quad \text{(conjugate perimeter)} \\
 \sigma = 4 \pi \ell_{\rm Planck}^2 \gamma \beta   \quad \text{(unit conjugate area)} \quad & \Leftrightarrow \quad \psi = 4 \pi \ell_{\rm Planck} \gamma \theta  \quad \text{(unit conjugate perimeter)} \\
\end{aligned}
\end{equation}}

\noindent  {Note that in the last line above, in order to anticipate the final result in the correct units, and to fix the proportionality factor in Eq. (\ref{ZT2}), we set:}
 {
\begin{equation}
\psi = 4 \pi \ell_{\rm Planck} \gamma \theta , \label{reparam}
\end{equation}}
\noindent  {where $\gamma$ is the Barbero-Immirzi parameter \cite{RovBOOK07}, \cite{Vya22}.}

\medskip

\noindent {The ensemble perimeter is then related to the partition function, Eq. (\ref{ZT2}), by:}
 {
\begin{equation}
\mathcal{C}_Q  = -{\partial \ln Z_T (\theta,N_{\rm h})\over \partial \theta}.\label{SMperim}
\end{equation}}
\noindent  {Therefore:}
 {
\begin{equation}
\left ( {1 \over 4 \pi \ell_{\rm Planck} \gamma N_{\rm h}} \right ) 
\mathcal{C}_Q = 
 - {\partial \ln z (\psi) \over \partial \psi} \equiv u, \label{SMperim2}
\end{equation} }
\noindent  {so that
\begin{equation}
\mathcal{C}_Q =  4 \pi \ell_{\rm Planck} \gamma N_{\rm h} u.
 \label{SMperim3}
\end{equation} }
\noindent  {By inverting the differential equation (Eq. \ref{SMperim2}), a solution for $\psi$ implies the functional dependence:}
  {
\begin{equation}
\psi = \psi(u) ~~~ ({\rm implying}~\Rightarrow~~~ \theta = \theta(u)).
\end{equation} }
\noindent  {We also set:
\begin{equation}
\mathcal{C}_Q =  N_{\rm h} f ~~~\Longrightarrow f \equiv 4 \pi \ell_{\rm Planck} \gamma u = {\psi(u) \over \theta} u.
 \label{SMperim4}
\end{equation} }
\noindent  {The entropy is formally given by \cite{Kra97}, \cite{Loc12} :
\begin{equation}
 S_Q = \ln Z_T (\theta,N_{\rm h}) + \theta \mathcal{C}_Q (\theta,N_{\rm h}),
 \label{entropy1}
\end{equation}}
\noindent  {or, using previous equations:
\begin{equation}
 S_Q = N_{\rm h} \left \{
 \ln z [\psi(u)]  + u \psi(u)
 \right \} .
 \label{entropy2}
\end{equation}}

 {The next step is to maximize $S_Q$ with respect to $N_{\rm h}$. However, from this point on, the derivation is, mathematically (for a reduced dimension), exactly as in Ref.  \cite{Loc12} (c.f. Eqs. (13) to (20) in their paper), which are here omitted. The final result is:}
 {
\begin{equation}
 S_Q = \psi(u_0) {\mathcal{C}_Q \over 4 \pi \ell_{\rm Planck} \gamma  },
 \label{entropySQ}
\end{equation}}
\noindent  {for some constant, $u_0$. We set the Barbero-Immirzi parameter as: $\gamma = \psi(u_0)/\pi$. Notice that in $2$ spatial dimensions, $\ell_{\rm Planck} = \hbar G$ \cite{RovBOOK07}. Therefore, Eq. (\ref{entropySQ}) leads to the same form as $S_{\rm BTZ}$ (Eq. \ref{BTZ_Entropy}), by taking the ensemble perimeter as $\mathcal{C}_Q = 2 \pi r_+$.}

\hspace{1cm}

{\sl \underline{Comments}.}  {The ensemble average of a perimeter within a region $Q$ of $\Omega$ (c.f. Eq. \ref{ENSAV}), $\mathcal{C}_Q$, corresponds in our framework to the event horizon of a BTZ black hole by taking the puncturing of relaxing TSN edges in the avalanche cycles to the perimeter representing the physical horizon. Indeed, as we established this framework for an arbitrary $\mathcal{C}_Q$, we can adequate it to a value representing any macroscopic horizon given in LQG, but the interpretation is that {\sl  it would arise only from  states which have intrinsically evolved through gauge non-invariant (GNI) avalanches}. }

\medskip

 {That is, the association between the perimeter of a BTZ black hole in our framework and the one given in (2+1)-dim. LQG \cite{Fro13}, \cite{RovVidBOOK20}, i.e.:}

 {
\begin{equation}
\mathcal{C}_Q ~~\Leftrightarrow ~~ L_{\rm LQG} = 8\pi \ell_{\rm Planck} \gamma \sum_i \sqrt {(j_i(j_i+1)}; ~~~ (j_i = \bar{c}_i/2), \label{CORRESP}
\end{equation}}

\noindent  {is {\sl not} arbitrary for any set of spins $j_i$. Only perimeter ensemble averages $\mathcal{C}_Q$'s resulting from SOC TSN microstates will correspond to the total length of given LQG BTZ horizons, which is to be understood when reading the correspondence given in Eq. (\ref{CORRESP}). We have not proven whether the correspondence above is one-to-one, in the sense that there could be instances in which a set of $j_i$'s would not result in SOC dynamics. In such case, there would be a BTZ horizon in LQG but not in our framework. We leave the proof of this later problem to a future work.}

\medskip

 {As a brief summary of our formalism, it is always possible to find SOC TSN microstates that will correspond to a LQG BTZ horizon. Any SOC path in which a set of color (spin) configurations, $Q = \{ \mathsf{A} \}$ (i.e., a region in $\mathcal{M}$) belongs to, will serve. Conversely, one can always construct some initial set $(\mathbf{a},Q_0) \in \Omega$, so that $Q$ is the result of the map $\mathcal{F}(\mathbf{a},Q_0)$, acted on for sufficiently large $\tau$ in the SOC dynamics, giving, therefore, a path of the SOC dynamics ($Q$ is $\mathcal{F}$-invariant; c.f. Prop. 2, and Eq. \ref{DYNSYS}). Assuming ergodicity (Eq. \ref{ERGO}) on the stability region, this would provide the perimeter estimate for the ensemble (Eq. \ref{ENSAV}). Finally, $\theta(u_0)$ in Eq. (\ref{ENTROPYperim}) would be adjusted to approach the BTZ entropy relation, Eq. (\ref{BTZ_Entropy}).}

\medskip

{ {
\section{Discussion and Conclusions}\label{SEC_conclusions}}

 {We summarize our main results and discuss some additional points of interest:}

 {
\begin{enumerate}
\item{{\sl An effort to formalize the microdynamics in LQG models exhibiting SOC using ergodic concepts.} To our knowledge, this paper introduces, for the first time, a mathematical formalism based on ergodic concepts for the propagation rules of a frozen TSN and establishes a basic starting point for extending these principles to quantum spin networks in broader contexts.}
\item{{\sl A contribution to the collection of methods that converge to the same perimeter-entropy form.} Although preliminary, our result offers an independent derivation of the BTZ black hole entropy using ergodic concepts, adding to the ``universality'' issue \cite{Car05}, \cite{Car07}. This raises an intriguing question: why was this result achievable, and how does it relate to previous approaches? We briefly mention the following possibility:}
 \subitem{-  In Ref. \cite{Loc12}, it is argued that any partition function with a certain form, representing in a sense ``noninteracting constituents'' making up the horizon, would lead generically to the Bekenstein-Hawking relation. Our SOC TSN partition function over a set of legal sequences of avalanches, Eq. (\ref{ZT2}), does have a form in agreement with \cite{Loc12}. Within our framework, the origin of these ``noninteracting constituents'' can be interpreted as follows. The dual triangulation of the finite TSN could be seen as embedded in a very large reservoir (semi-classical spacetime). Vertices at the boundary disturb the reservoir only infinitesimally. The activation of a vertex is an stochastic disturbance from the reservoir. Other TSN members of the ensemble are connected via the reservoir and are not directly affected, so the ensemble represent granularities of spacetime undergoing independent avalanches, which can be taken as the fundamental noninteracting constituents over the domains of stability.}
\item{{\sl A proposal for a different microscopic origin of the BTZ entropy.} Our work proposes an alternative approach to deriving the BTZ entropy law, utilizing a state-counting method that depends exclusively on the presence of SOC, that is, from the excitation-relaxation spin dynamics within the avalanche cycle. These degrees of freedom do not reside in a pre-established boundary in the TSN but emerge dynamically from the stability domains connecting gauge non-invariant avalanches. We interpret the puncturing of TSN edges involved in the avalanche as an ensemble perimeter over these stability domains. Only in the final step of our derivation is this perimeter associated with that of an isolated (2+1)-dimensional black hole horizon. We establish a connection between the resulting SOC TSN entropy and the entropy of a BTZ black hole through a potential function in our formalism, showing that both entropies converge to the same form. An interesting possibility would be to connect our results with the ``stretched horizon'' constraint as suggested in Ref. \cite{Car07}. In this sense, a correspondence between our ensemble perimeter and a ``dynamical horizon'' \cite{Boo05} would be an interesting step to verify the robustness of our framework.}
\item{{{\sl Leading term for the classical BTZ entropy.} It is important to note that we considered a simplification in our derivations, by absorbing the ``excitation entropy'' in the potential function (Eq. \ref{POTnew}) used in defining the partition function (Eq. \ref{POTTDF}). Consequently, our results represent the leading term of the perimeter-entropy. We leave for a future work an analysis of whether the excitation entropy can be associated with further LQG corrections to the entropy \cite{Mei2004}. Indeed,  the discrete nature of spacetime in LQG could regularize high-energy behavior, preventing singularities and leading to classically to regular black holes \cite{Bue2021}. This scenario often results in logarithmic corrections to the entropy, as seen in various studies  \cite{Ma2014} \cite{Ali2022}, \cite{Akh2023}. }}
\end{enumerate}
}

\medskip

{We briefly comment on the possibility of extending our work to the full $4$-dimensional Lorentzian case (spin foam models)  \cite{Per2013}. Notably, to the best of our knowledge, there are no specific studies investigating SOC in numerical spin foam models. This gap in the literature makes it challenging to directly extend the SOC conjectures from the $(2+1)$-dimensional frozen TSN case (where SOC occurs under certain conditions) to the full $4$-dimensional Lorentzian LQG case (where the presence of SOC is unknown). However, if SOC is found to be present in the full case, it would then be a matter of mapping its conditions and the corresponding spin foam dynamics in terms of the dynamical systems framework proposed here. }

\medskip

{However, the transition from a frozen TSN model with spin propagation rules to a full LQG framework, e.g., represented by spin foam dynamics, represents a significant increase in conceptual complexity. Additionally, it remains uncertain how to effectively model gauge non-invariant avalanches as an internal excitation-relaxation process within this more intricate dynamical framework. Spin foams represent histories of spin networks, where vertices, edges, and faces of a $2$-complex are labeled by spins and intertwiners, capturing how spin networks evolve over time. This transition involves solving complex path integrals over all possible spin foam configurations, which is conceptually intricate, especially when formulating it as a dynamical system. The need to account for both the quantum superposition of geometries and the intricate vertex amplitudes adds layers of complexity absent in the simpler TSN models. Thus, while it is not currently feasible to expand our application immediately to the 4D case, it could inform future studies within the full LQG framework.}

\medskip

 {We conclude with considerations regarding observational and experimental predictions and tests of our proposal. Given the dimensionally reduced nature of the SOC TSN model and the derived entropy based on avalanches in gauge non-invariant spin microdynamics, initial steps should focus on preliminary checks for the overall consistency of the rationale.}
 {
\begin{itemize}
\item{One possibility is to perform simulations to directly measure the asymptotic SOC TSN entropy. Although previous numerical studies primarily focused on verifying SOC \cite{Ans08}, \cite{Che08}, \cite{Dan21-CQG}, similar codes could be extended and rerun to compute the SOC TSN entropy. Note that the {\sl size of the avalanche} is formally given by the number of updates in a cycle (i.e., color additions to the edges of vertices, in which repeated vertices are counted). This provides a counting algorithm that is essentially a measure of the SOC TSN entropy, which however should be computed after a sufficiently large number of cycles and for a large number of asymptotic states. A systematic scaling involving the resulting  entropy and the ensemble perimeter for various initial conditions, TSN sizes, and models \cite{Dan21-CQG} would then indicate a consistency with our results.}
\item{ Alternatively, one could attempt to devise experimental setups that simulate gravitational phenomena in laboratory conditions, known as analog gravity systems, which can mimic aspects of black hole physics \cite{Bar11}. {Some proposals devise quantum mechanical systems that could be affected by quantum gravity theories. For instance, the effects of a minimum observable length and maximum observable momentum on the transition rate of ultra-cold neutrons in a gravitational field has been investigated, utilizing the Generalized Uncertainty Principle \cite{Ped2011}. Additionally, in Ref. \cite{Kho2018}, it has been proposed an opto-mechanical experimental setup to explore the effects of polymer quantization \cite{Ash2003}. Their approach demonstrates the feasibility of using high-intensity optical pulses inside an optical cavity to probe quantum gravitational phenomena. Such controlled experiments} to replicate the effects of propagation rules, excitation-relaxation microdynamics with avalanches, etc., would be necessary, but currently pose significant challenges.}
 \end{itemize}
 }


\hspace{2cm}

\medskip
\textbf{Acknowledgements} \par 
{CCD thanks the referees for their comments and suggestions, which have greatly improved the quality and clarity of this paper.}
This study was financed in part by the Coordena\c c\~ao de Aperfei\c coamento de Pessoal de N\'{\i}vel Superior - Brasil (CAPES) - Finance Code 001, the Programa Institucional de Internacionaliza\c c\~ao (PrInt – CAPES), and the Brazilian Space Agency (AEB) for the funding (PO 20VB.0009).


\medskip

%
\bibliographystyle{MSP}
\bibliography{CCDANTAS2}


\end{document}